\documentclass[aip,amsfonts,amsmath,amssymb,reprint]{revtex4-2}
\usepackage{graphicx}% Include figure files
\usepackage{dcolumn}% Align table columns on decimal point
\usepackage{bm}% bold math
\usepackage{xcolor}
\usepackage[english]{babel}
\usepackage[utf8]{inputenc}
\usepackage[T1]{fontenc}
\usepackage{mathptmx}
\usepackage{tabularx,booktabs}
\usepackage{soul}
\usepackage{color}

\setstcolor{red}

\begin{document}

\preprint{----}

\title{Betweenness centrality of teams in social networks}

\author{Jongshin Lee}
%  \altaffiliation[Also at ]{Physics Department, XYZ University.}%Lines break automatically or can be forced with \\
\author{Yongsun Lee}%
\author{Soo Min Oh}%
\author{B. Kahng}%
\email{bkahng@snu.ac.kr}
\affiliation{ 
CCSS, CTP, Department of Physics and Astronomy, Seoul National University, Seoul 08826, Korea
}%

\date{\today}% It is always \today, today,

\begin{abstract}
  Betweenness centrality (BC) was proposed as an indicator of the extent of an individual's influence in a social network. It is measured by counting how many times a vertex (i.e., an individual) appears on all the shortest paths between pairs of vertices. A question naturally arises as to how the influence of a team or group in a social network can be measured. Here, we propose a method of measuring this influence on a bipartite graph comprising vertices (individuals) and hyperedges (teams). When the hyperedge size varies, the number of shortest paths between two vertices in a hypergraph can be larger than that in a binary graph. Thus, the power-law behavior of the team BC distribution breaks down in scale-free hypergraphs. However, when the weight of each hyperedge, for example, the performance per team member, is counted, the team BC distribution is found to exhibit power-law behavior. We find that a team with a widely connected member is highly influential.
\end{abstract}

\maketitle

\begin{quotation} %lead paragraph
  A graph consisting of pairwise interactions was useful for understanding emerging phenomena such as the formation of a giant community, and quantifying an individual's influence by betweenness centrality (BC). However, the BC is not useful to measure the influence of a team that is composed of more than two people. Here, we extend mesurement method of the BC in a graph to a hypergraph with higher-order interactions, which is not straightforward. We find that the team BC distribution in a scale-free binary hypergraph does not exhibit power-law behavior because the number of the shortest paths between every pair of nodes increases significantly. We find that a weight of hyperedge, the performance per team member is appropriate to obtain the power-law behavior of the team BC distribution. Interestingly, it reveals that the team with high performance per team member does not have large team BC value, but the team with a leader with large degree has large team BC. Thus, our result supports the statement that connectivity is more crucial than performance for success.
\end{quotation}

Complex systems comprising many elements and their interactions can be represented by graphs for simplicity. A graph is composed of vertices and edges, which represent elements and pairwise interactions among the elements, respectively~\cite{complexnet_keypaper_3, complexnet_keypaper_4, complexnet_keypaper_5, complexnet_keypaper_6}. A hypergraph is a generalization of a graph in which interacting elements are not limited to pairs of vertices; that is, more than two elements can interact, for instance, coauthors of a paper, players on a sports team, or proteins in a protein complex in biological networks. These interacting teams are represented by hyperedges in hypergraph representation~\cite{berge1984hypergraphs,voloshin2009introduction,PhysRevE.74.056111,PhysRevE.77.046104,PhysRevE.77.066106,10.1371/journal.pcbi.1000385,Newman2009Hypergraph,PhysRevE.80.036118,10.1145/1873951.1874005,Taramasco2010,Liu2018,jhun2019simplicial,carletti2020random,de2020social,Alvarez-Rodriguez2021,FerrazdeArruda2021}. 

Betweenness centrality (BC) was proposed as a measure of the influence of a vertex (individual) in a social network. This quantity is measured by counting the number of times a vertex appears on all the shortest paths between pairs of vertices in a graph. The vertex BC ($v$-BC) of vertex $v_i$ [$B(v_i)$] is obtained as 
%\begin{eqnarray}
%B(v_i)=\sum_{s\neq v \neq t}\frac{\sigma_{st}(v_i)}{\sigma_{st}}
$B(v_i)=\sum_{s\neq v_i \neq t}{\sigma_{st}(v_i)}/{\sigma_{st}}$,
%\label{eq:one}\,,
%\end{eqnarray}
where $\sigma_{st}$ is the number of shortest paths between vertices $s$ and $t$, and $\sigma_{st}(v_i)$ represents the number of shortest paths passing through vertex $v_i$~\cite{freeman1977set,Freeman1978BC}. 
BC has been used to describe diverse phenomena such as the load of each router on the Internet~\cite{erramilli1996experimental, Goh2001UniversalLoadDist, tadic2002packet, tadic2004information, tadic2004traffic, tadic2005search, dodds2003information,Kirkley2018,Estrada2018CentralitiesInSC,Zaoli2021}. Interestingly, the BC of each vertex has a power-law distribution in scale-free (SF) networks, and the exponent differs from that of the degree distribution~\cite{Goh2002ClassificationScaleFree, goh2005load}. The edge BC ($e$-BC) of edge $e_i$ is defined similarly to $v$-BC as the fraction of shortest paths passing through edge $e_i$ among all the shortest paths between pairs of vertices~\cite{Girvan7821,Grady2012}.

Here, we are interested in the BC of a group or team composed of more than two people, for instance, the influence of a research group comprising coauthors of a paper in a coauthorship network or the importance of the role a protein complex plays in signal transduction in a protein interaction network~\cite{signal}. However, it is not obvious whether $v$-BC or $e$-BC is useful for this purpose. For instance, when three researchers form a team, a 3-clique represents the linkage of the group; however, this clique may be interpreted as the representation of three groups composed of pairs of vertices. Thus, it is challenging to introduce an appropriate quantity that measures the influence of a higher-order interacting object such as a group or team. Here, we propose the use of the hyperedge BC ($h$-BC) in hypergraphs for this purpose. 

\begin{figure*}[t]
\includegraphics[scale=0.65]{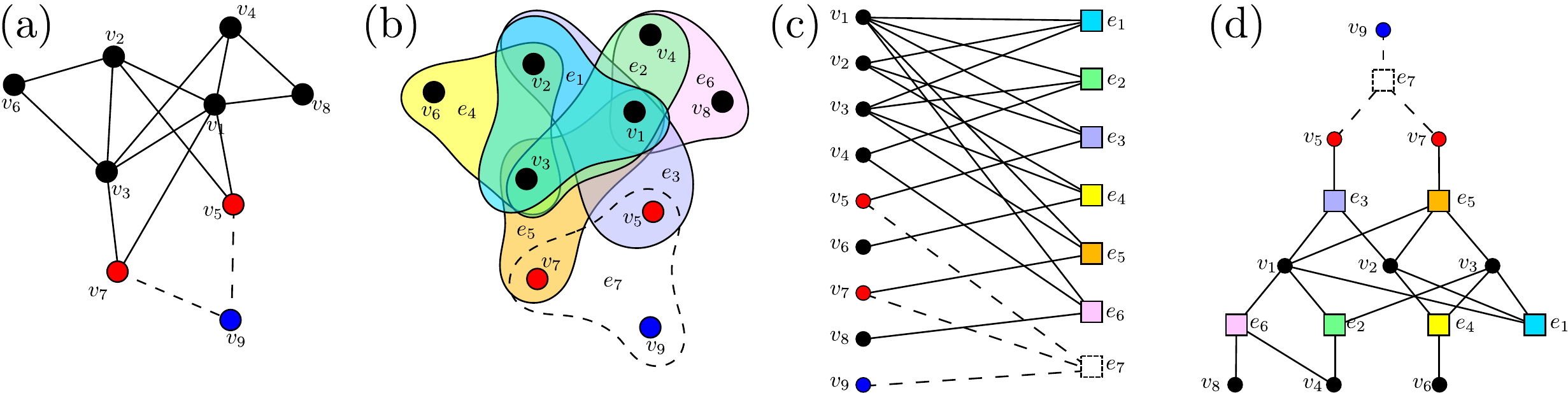}
\caption{\label{fig:HGtoBipartite} 
Diagram illustrating the measurement of $h$-BC. From graph (a), hypergraph (b) is constructed. The hypergraph is represented by a bipartite graph (c) comprising vertices and hyperedges. Next, all the shortest paths (d) between pairs of vertices are identified using a given rule. Then, the BC of each vertex and hyperedge is determined.}
\end{figure*}

To obtain the $h$-BC in a hypergraph, we use a bipartite graph. For instance, in Fig.~\ref{fig:HGtoBipartite}, two adjacent vertices are connected by one or more hyperedges with higher-order interactions. Then, the path between any two vertices on the hypergraph is defined as an alternating sequence of vertices and adjacent hyperedges between the two vertices. The paths with the smallest hopping steps between vertex and hyperedge sets are the shortest. The $h$-BC is defined similarly to the $v$-BC along the shortest paths in a graph. 

We performed the computation using the algorithm introduced by Brandes \cite{brandes2001faster}. Briefly, a shortest-path tree is constructed by choosing a vertex and then stacking the BC values while returning. Repeating this process for all the starting points yields the entire BC. Fig.~\ref{fig:HGtoBipartite} illustrates the entire process of measuring the $v$-BCs in the bipartite graph. Note that hyperedges are regarded as vertices in the bipartite representation. 

\begin{figure*}[th!]
	\centering
\includegraphics[scale=0.40]{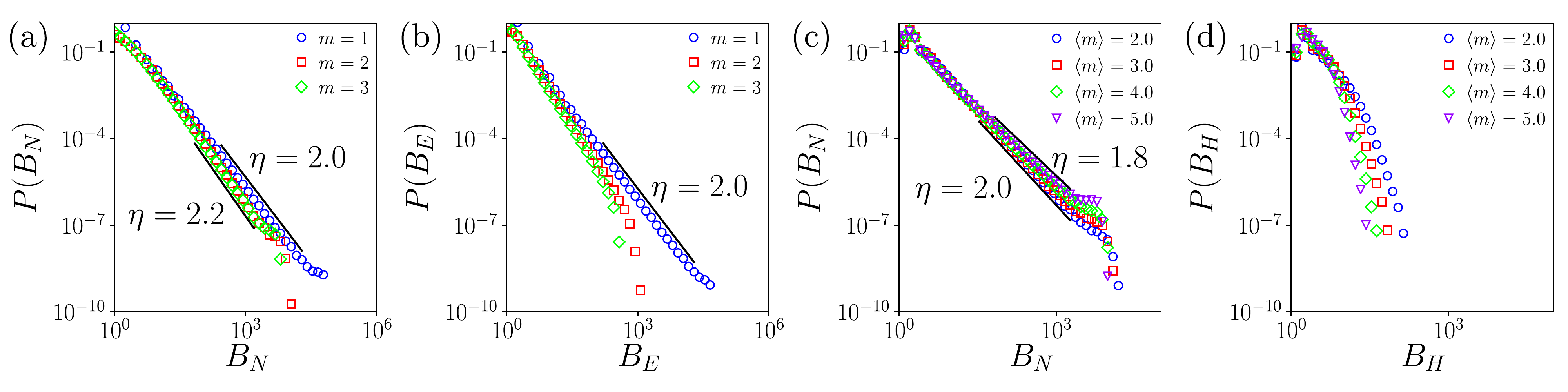}
\caption{\label{fig:fig2} (a) $v$-BC distribution and (b) $e$-BC distribution for the BA graph for several $m$ values. (c) $v$-BC distribution and (d) $e$-BC distribution for the BA hypergraph with several mean hyperedges $\langle m \rangle$. Numerical data are obtained for the system size $N=10^5$ and averaged over $10^2$ ensembles for each data point.}
\end{figure*} 

We first consider one graph and one hypergraph as minimum models and obtain the distributions of $v$-BC, $e$-BC and $h$-BC. The two minimal model networks are as follows: 
\begin{itemize} 
	\item[(i)] The model is the graph composed of vertices and edges~\cite{Barabasi1999BAModel} proposed by Barab\'asi and Albert(the BA model). The BA model evolves as follows. (a) The system initially contains $m$ isolated vertices. (b) At each time step, a vertex is added to the system and is connected to $m$ existing vertices selected according to the so-called preferential attachment rule, that is, a vertex $v_i$ with degree $k_i$ is selected with probability $k_i/\sum_j k_j$. This process is repeated $N-m$ times, and an SF graph with $N$ vertices is constructed. The degree distribution of an SF graph follows the power law $P_d(k)\sim k^{-\gamma}$. For the BA graph, the exponent $\gamma=3$ is obtained. 
	\item[(ii)] BA hypergraph: step (a) is the same as that for the BA model. In step (b), the number of target vertices $m$ is taken from the Poisson distribution with given mean values $\langle m \rangle$. A new vertex and the $m$ selected target vertices are regarded as the elements of a hyperedge. The preferential attachment rule is modified such that each vertex $v_i$ among the $m$ vertices is selected with a probability proportional to the hypergraph degree $f_i$, that is, the number of hyperedges vertex $v_i$ belongs to, which is given as ${f_i}/{\sum_j f_j}$. 
	%\item[(iii)] BA-II hypergraph: the number $m$ of target vertices for the rule (b) is fixed. 
\end{itemize}
%Second, we will consider an empirical hypergraph for the study of $h$-BC distribution.  

For this graph and hypergraph, we investigate the $v$-, $e$-, and $h$-BC distributions for different $m$ or $\langle m \rangle$ values. Fig.~\ref{fig:fig2}(a) and \ref{fig:fig2}(b) show the results for the BA graph, and Fig.~\ref{fig:fig2}(c) and \ref{fig:fig2}(d) shows those for the BA hypergraph. The $v$-BC distributions for the graph and the hypergraph are shown in \ref{fig:fig2}(a) and \ref{fig:fig2}(c), respectively. The corresponding $e$- and $h$-BC distributions are shown in Fig.~\ref{fig:fig2}(b) and \ref{fig:fig2}(d), respectively. 

We first consider the $v$-BC distribution. (i) For the BA graph, it exhibits power-law behavior as $P(B_N)\sim B_N^{-\eta}$ with exponents $\eta\approx 2.01$ for $m=1$ and $\eta \approx 2.17$ for $m=2$ and $3$. When $m=1$, the graph is a tree, and the exponent $\eta$ was analytically obtained as $\eta=2$~\cite{Goh2002ClassificationScaleFree}. However, when $m \ge 2$, the graph has loops, and the analytic solution for this case has not been reported. 

%For BA-I hypergraph (ii), the distributions of $v$-BCs for different $m$ values also exhibit power-law behaviors as shown in Fig.~\ref{fig:fig2}(c). The exponent $\eta$ for different $m$ seems to depend on $m$ weakly: $\eta \approx 2.10$ for $m=2$ is larger than $\eta\approx 1.80$ for $m=4$. Thus $\eta$ values for smaller $m$ are larger than $\eta$ for larger $m$. However for the case (i), the exponent $\eta$ for smaller $m$ is smaller than the one for larger $m$. This difference originates from the point: As $m$ is raised, each hyperedge contains a larger number of vertices. This induces more shortest pathways. Recall that a pathway between two vertices is an alternating sequence of vertices and hyperedges. Even though we deal with the hypergraph, the shortest pathways do not significantly differ from those of the graph case. However, the number of shortest pathways passing by each vertex becomes more in a rich-get-richer way for hypergraphs. Hence, the $v$-BCes of the vertices with large hyperedge degree are increased by relatively a large amount compared with those of vertices with small hyperedge degree. This leads to the smaller $\eta$ value of $v$-BC for larger $m$ case. Furthermore, the $v$-BC distribution exhibits a small bump in the tail part for the cases $m=3$ and 4. 

The BA hypergraph is designed to model a coauthorship hypergraph consisting of authors and papers. An author (paper) is regarded as a vertex (hyperedge). The number of authors of a paper, which corresponds to the size of a hyperedge, follows a Poisson distribution with mean degree $\langle m \rangle \approx 2.5$. Here, we control the mean degree $\langle m \rangle$ of the hypergraph. The $v$-BC distributions for different $\langle m \rangle$ values exhibit power-law behavior with an exponent of $\eta\approx 2.0$ for $\langle m \rangle=2$, which gradually decreases with increasing $\langle m \rangle$ (Fig.~\ref{fig:fig2}(c)). Overall, the $v$-BC distributions of the graph and hypergraph exhibit power-law behavior. %However, it is likely that the BC exponent slightly decreases when a more realistic structure is applied. 

\begin{figure*}[!t]
\includegraphics[scale=0.40]{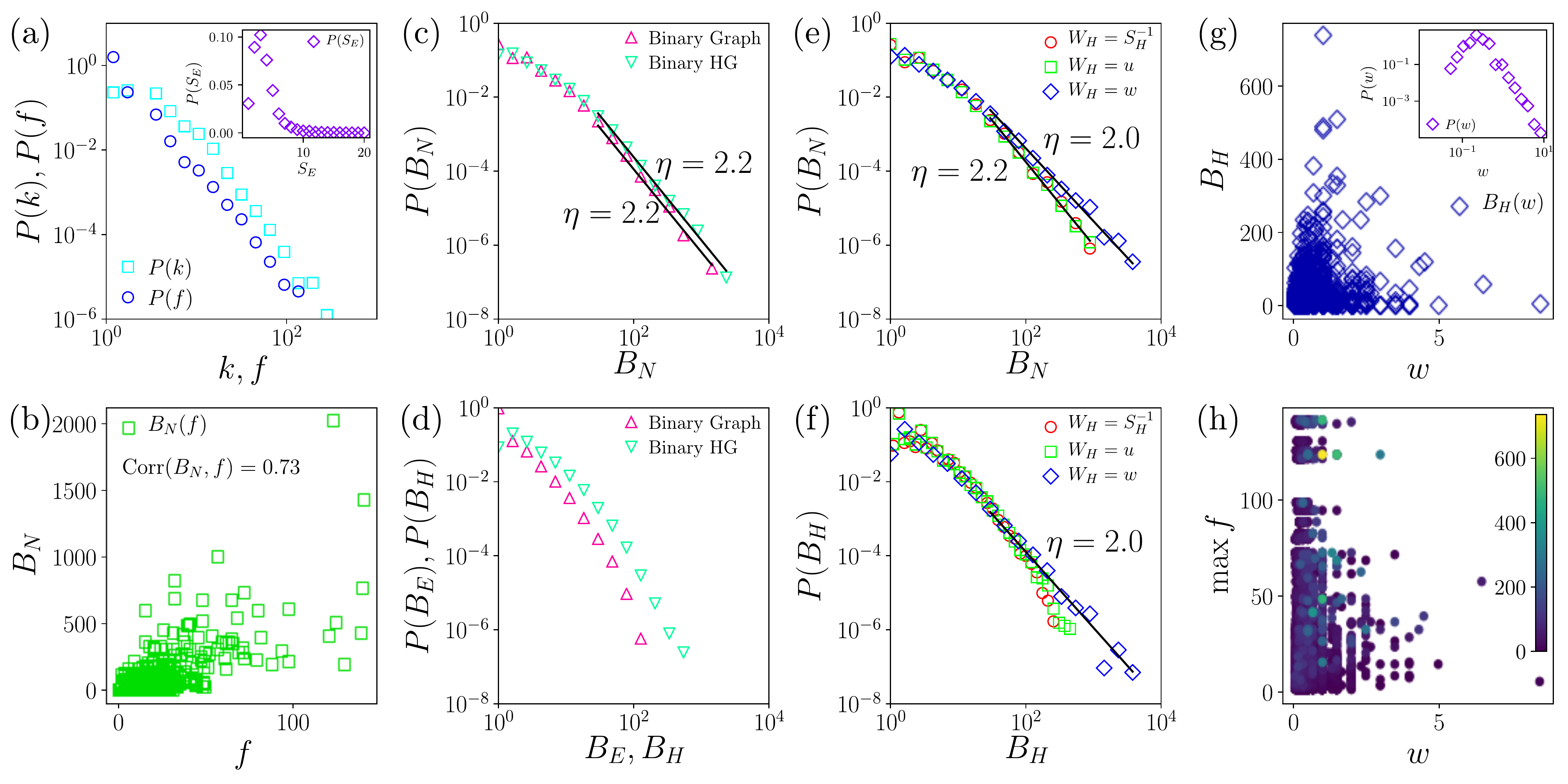}
\caption{\label{fig:fig3} (a), Degree distributions of coauthorship graph ($\square$) and hypergraph ($\bigcirc$), which exhibit power-law behavior with slopes $-2.9$ ($\square$) and $-2.7$ ($\bigcirc$). Inset: The distribution of hyperedge sizes $S_E$ (i.e., the number of coauthors) for the coauthorship hypergraph on a linear scale. The distribution follows a Poisson distribution with mean value $2.5$. (b), Plot of $B_N(f)$ versus hyperedge degree $f$. It is likely that a vertex with a larger hyperedge has a larger $B_N$. (c), (d), (e), and (f), Various types of BC distributions $P(B_X)$, where $X=N, E$, or $H$, which represent nodes, edges, and hyperedges, respectively, for the coauthorship graph and hypergraph. (c) and (d) are those of the graph and hypergraph with binary edges, respectively. (e) and (f) are those of $u$-weighted and $w$-weighted hypergraphs, respectively. Legend: $W_H=S_H^{-1}$, $W_H=u$, and $W_H=w$ represent the type of weight, specifically, the inverse of the number of group members, the number of papers, and the number of papers per group member, respectively. The data are logarithmically binned. Slopes of the BC distributions are measured asymptotically in the tails. (g), Plot of $B_H$ as a function of weight $w$. This plot indicates that hyperedges with larger weight $w$ do not have larger $B_H$. Inset: The distribution of the weight $w$ of the coauthorship hypergraph. (h), Three-dimensional plot of $h$-BC (team BC) of each hyperedge as a function of weight $w$ and the maximum $f$ value of the hyperedge degree of each element belonging to the given hyperedge. The BC of a large team is more likely to appear on a hyperedge with a large $f$ value, which represents a team with a super leader.}
\end{figure*}

Next, we consider the distributions of $e$-BC and $h$-BC in Fig.~\ref{fig:fig2}(b) and \ref{fig:fig2}(d), respectively. For the BA tree graph with $m=1$ in \ref{fig:fig2}(b), the $e$-BC distribution is very similar to the $v$-BC distribution. For $m \ge 2$, the $e$-BC distribution exhibits power-law behavior for small $B_E$; however, it shows exponential decay for large $B_E$. For the BA hypergraph with different $\langle m \rangle$ values, the $h$-BC distributions decay more rapidly as the mean hyperedge size $\langle m \rangle$ is increased. The reason is as follows. As $\langle m \rangle$ is increased, the total number of shortest paths between pairs of vertices shows a faster than linear increase with respect to $\langle m \rangle$. Thus, hyperedges may be more likely to appear on the shortest paths. More shortest paths can be generated between two given vertices in hypergraphs. Consequently, the hyperedge has a somewhat smaller $h$-BC value, because $h$-BC is the inverse of the total number of shortest paths between two given vertices. This behavior is more apparent when $\langle m \rangle$ is large.   

%\begin{figure}[t]
%	\includegraphics[scale=0.40]{fig4.pdf}
%	\caption{\label{fig:CoauthorshipBC} Various types of BC distributions $P(B_X)$ with $X=N, E$, or $H$ representing node, edge, and hyperedge, respectively, for the coauthorship graph and hypergraph. (a) and (c) are for graph and hypergraph with binary edges, respectively. (b) and (d) are for $u$-weighted and $w$-weighted hypergraph, respectively. Legends: $W_H=S_H^{-1}$, $W_H=u$, and $W_H=w$ represent the types of weight: the inverse of the number of group members, the number of papers, and the number of papers per group member, respectively. The data are logarithmically binned. Slopes of the BC distributions are measured asymptotically in the tail parts.}
%\end{figure}

We used a coauthorship dataset~\cite{yslee2020} consisting of authors and papers in the field of network science published from June 1998 to the end of 2017 that cite two pioneering papers, those on the Watts--Strogatz model of small-world networks and the Barab\'asi--Albert model of SF networks, and highly cited early review papers~\cite{Watts1998SmallWorld,Barabasi1999BAModel,complexnet_keypaper_3,complexnet_keypaper_4,complexnet_keypaper_5,BOCCALETTI2006175}. The dataset contains 32,016 distinct authors and 21,653 papers~\cite{Newman404,patania2017shape}. For convenience, we consider the BC distribution on the giant hypergraph. In the graph representation, authors are connected if they have written at least one paper together. Here, we constructed a coauthorship hypergraph by regarding papers as hyperedges. Fig.~\ref{fig:fig3}(a) shows the structural properties (the graph degree and hypergraph degree distributions) of the giant cluster in the coauthorship network in the graph and hypergraph representations. These degree distributions exhibit power-law behavior with $P_g(k)\sim k^{-\gamma}$ and $P_h(k)\sim k^{-\delta}$, respectively. The inset of Fig.~\ref{fig:fig3}(a) shows the distribution of the hyperedge size, i.e., the number of coauthors of each paper. It seems to fit the Poisson distribution with a mean value $\langle m \rangle=2.5$. 

The BC distributions of the coauthorship graph and hypergraph are investigated by measuring the $v$-BC, $e$-BC, and $h$-BC distributions. The $v$-BC distributions of the coauthorship graph and hypergraph seem to exhibit power-law behavior in the tail region with exponent $\eta\approx 2.2$, as shown in Fig.~\ref{fig:fig3}(c). Note that the vertices with large hyperedge degrees have large $v$-BC values, as shown in Fig.~\ref{fig:fig3}(b). The reason is probably that the vertices with large graph degrees have large $v$-BC values. The $e$-BC and $h$-BC distributions follow heavy-tailed distributions overall, as shown in Fig.~\ref{fig:fig3}(d).

\begin{figure*}[t]
\begin{center}
\includegraphics[scale=0.75]{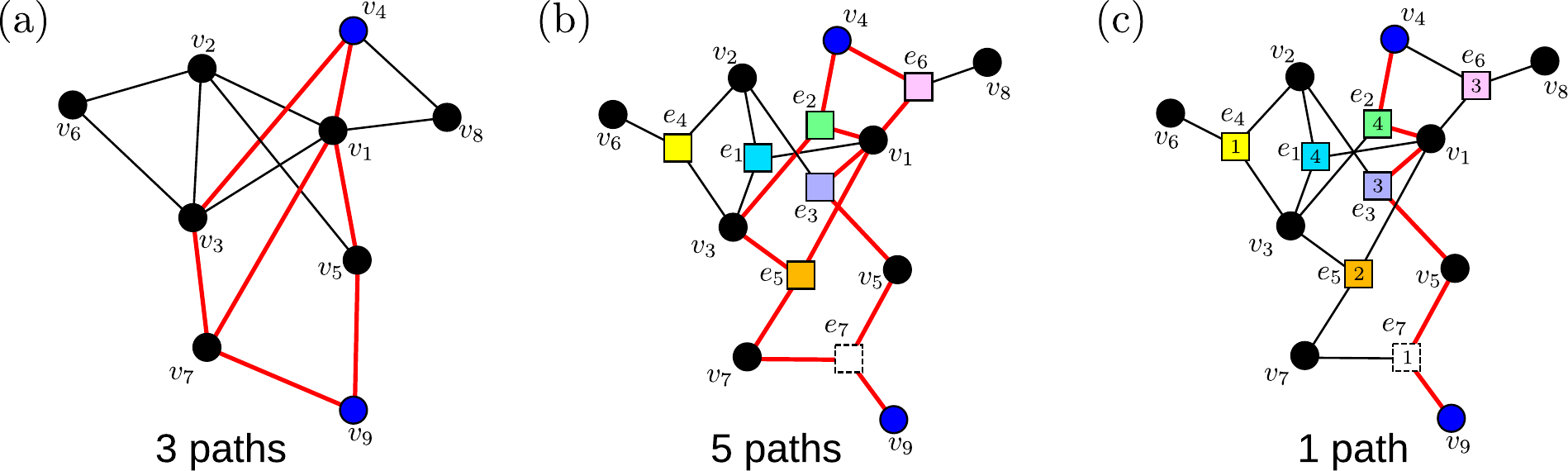}
\end{center}
\caption{\label{fig:fig4} 
Illustration of the differences in shortest paths between a given pair of vertices, $v_4$ and $v_9$ (blue nodes) (a) on a graph, (b) on a binary hypergraph, and (c) on a weighted hypergraph. Vertices ($\bigcirc$) are denoted as $\{v_i\}$. Hyperedges ($\square$) are denoted as $\{e_i\}$. Numbers in boxes represent the weights of hyperedges. Lines represent interactions. Thick lines represent the shortest paths between vertices $v_4$ and $v_9$. There are three paths in (a), five paths in (b), and one path in (c).}
\end{figure*}

We note that each hyperedge in the coauthorship hypergraph represents the team of coauthors who wrote each paper. The empirical dataset shows how many papers each research group has published. Using this information, we construct a weighted coauthorship hypergraph in which each hyperedge is assigned two types of weights. One is the number of papers published by a team, which is expressed as $u_i$, where $i$ is a research group index. We regard this weight as representing the performance of team $i$. The other weight is the number of papers divided by the number of coauthors, which is denoted as $w_i$. This weight is interpreted as the performance per group member. %\bnew{\sout{In other words, it is a number that quantifies how much each member contributes to the team's performance.}}

We first consider how to identify the shortest paths between a pair of vertices and then how to measure the BC of each vertex and hyperedge. Let us consider the weight $u_i$. We introduce the cost of each hyperedge $i$ as $c_i\equiv 1/u_i$ and then find the shortest path(s) between two vertices $s$ and $t$ along which the total cost $\sum_{i\in \ell}c_i$ is minimum, where the index $i$ runs through all hyperedges along a shortest path $\ell$ between nodes $s$ and $t$. If there is one shortest path, the BC of each hyperedge $i\in \ell$ along the shortest path $\ell$ is set to $1$ for the pair of vertices $(s,t)$. However, if the number of shortest paths is larger than one, say $n$, a BC of $1/n$ is assigned to each hyperedge on the shortest paths $\{\ell_j \}$, where $j=1\cdots n$. This process is repeated for every pair of vertices $(s,t)$ in the coauthorship hypergraph. The BC values of each vertex and hyperedge are then the sum of all those BCs; the $v$-BC and $h$-BC distributions are indicated in Fig.~\ref{fig:fig3}(b) and \ref{fig:fig3}(d) by inverted triangles ($\bigtriangledown$). 
We also obtain the $v$-BC and $h$-BC for the weight $\{ w_i \}$ in a similar way. The data ($\Diamond$) in Fig.~\ref{fig:fig3}(e) and \ref{fig:fig3}(f) show the $v$-BC and $h$-BC distributions, respectively. Both of these BC distributions exhibit power-law behavior with exponent $\eta\approx 2.0$. The exponent value of $\approx 2.0$ may imply (but not guarantee) that there is usually one shortest path between two vertices. The weight $\{w_i\}$ enables us to select the most significant shortest path from among the multiple shortest paths generated in the hyperedge representation. We remark that although the $h$-BC distribution of the coauthorship network in binary hypergraph representation in Fig.~\ref{fig:fig3}(d) does not exhibit power-law behavior, the distribution in the weighted hypergraph representation, particularly that with weight $\{w_i\}$, does clearly show power-law behavior (Fig.~\ref{fig:fig3}(f)). The greatest heterogeneity among the four paths is achieved by the $w$-weighted hypergraph. %\rnew{\sout{Accordingly, the $w$-measure is appropriate to achieve the heterogeneity from the influences of each group.}} 

%We check if the tail part of the $h$-BC distribution based on the $w$-weight measure is contributed by the hyperedges with large $w_i$ values. However, their $w_i$ values are not so large as shown in Fig.~\ref{fig:fig3}(g). \bnew{\sout{Most teams have small $w_i$s. Since the distribution of $w$-weight is inhomogenous, it is appropriate to think differently depending on the density. We are interested in the high-performance teams, so we investigated the area after the density changes rapidly. The maximum value of BC appears at the point where $w$ begins to sparse.}} Counterintuitively, the tail part of the $h$-BC distribution is contributed by the hyperedges containing the nodes with large hypergraph degree $f_i$ (i.e., large $v$-BC values) in the bipartite representation as shown in Fig.~\ref{fig:fig3}(h). These nodes may be team leaders of each hyperedge.

We check whether the tail of the $h$-BC distribution based on the $w$-weight measure is contributed by the hyperedges with large $w_i$ values and find that their $w_i$ values are not very large, as shown in Fig.~\ref{fig:fig3}(g). 
%\bnew{\sout{Most teams have small $w_i$s. Since the distribution of $w$-weight is inhomogenous, it is appropriate to think differently depending on the density. We are interested in the high-performance teams, so we investigated the area after the density changes rapidly. The maximum value of BC appears at the point where $w$ begins to sparse.}} 
Counterintuitively, the tail of the $h$-BC distribution is contributed by the hyperedges containing the nodes with large hypergraph degree $f_i$ (i.e., large $v$-BC values) in the bipartite representation, as shown in Fig.~\ref{fig:fig3}(h). These nodes may represent the team leaders of each hyperedge.

%\bnew{Our analysis shows that the heterogeneity of the $h$-BC distribution is achieved when $w$ weights of each hyperedge are taken into account. The tail part of the power-law distribution of the team BC is mainly contributed by the hypereges next to the node with large hypergraph degree, that is, by a team leader with a large hypergraph degree.}  

%Since the optimal path is defined as a path that minimizes the overall cost in the weighted hypergraph, it is obvious that a different distribution appears because of the distorted the shortest-path itself. The reason for the decrease in the exponent is based on the characteristics of the data we used. Since a team that has existed for a long time would have written more papers over the entire period, the given weights are inhomogeneously distributed. Accordingly, BC distribution reflects this unevenness.}

In summary, we investigated the BC distributions of SF hypergraphs in various forms. Unlike the $e$-BCs in an SF graph, the $h$-BCs are meaningful because they represent the degrees of influence of teams or groups in social networks. By mapping a hypergraph to a bipartite network, we measured the BC distributions of vertices and hyperedges simultaneously. Counterintuitively, the $h$-BC distribution in an SF hypergraph does not exhibit power-law behavior. The reason is that the number of shortest paths between a given pair of vertices in a hypergraph becomes more than that in the graph, as shown in Fig.~\ref{fig:fig4}(a) and \ref{fig:fig4}(b), which reduces the contribution of each hyperedge to communication between a given pair of vertices. Therefore, it was challenging to design a BC measure that generates a heterogeneous $h$-BC distribution. To overcome this difficulty, we proposed a weighted hypergraph in which each hyperedge (team) is assigned a weight that is proportional to the activity per group member. Assuming that information spreads between two vertices along the paths that minimize the total cost, which is given as the sum of the inverse of the $w$-weights of each hyperedge along the path, we found that the number of shortest paths decreases to one, as shown in Fig.~\ref{fig:fig4}(c). Thus, the $h$-BC distribution exhibits power-law decay with exponent $\eta=2.0$. This BC measure is useful for quantifying team influence in a coauthorship hypergraph. As Barab\'asi stated in his book~\cite{barabasi2018formula}, ``Performance drives success, but when performance cannot be measured, networks drive success.'' Our results support the importance of networking in the transfer of information even for high-performing teams. We expect that the methodology developed using the coauthorship hypergraph will be useful in other hypergraphs. 

\begin{acknowledgments}
This research was supported by the NRF of Republic of Korea, Grant No.~NRF-2014R1A3A2069005 (BK).
%{\bf Author's contributions}: Y.L., D.L., and B.K. designed the research; Y.L. and J.L. performed numerical simulations; and Y.L., J.L., and S.M.O. set up the model \sout{and obtained analytic solutions}. B.K. conducted the research and wrote the first draft of the manuscript. All authors discussed results and edited the manuscript and approved the final version.\\ 
%{\bf Competing interests}: The authors declare that they have no competing interests. \\ 
% {\bf Data availability}: All data used in this work are available from the authors upon reasonable request.
\end{acknowledgments}

\section*{Data availability}
All data used in this work are available from the authors upon reasonable request.
% \nocite{*}
% \bibliography{bib}% Produces the bibliography via BibTeX.
%
\end{document}